\documentclass{PoS}

\def\Slash#1{\not\!\!#1}

\title{Analytical relation between confinement and chiral symmetry breaking in 
terms of Polyakov loop and Dirac eigenmodes in odd-number lattice QCD}

\ShortTitle{Analytical relation between confinement and 
chiral symmetry breaking in lattice QCD}

\author{\speaker{Hideo Suganuma}, Takahiro M. Doi \\
        Department of Physics \& Division of Physics and Astronomy, 
Graduate School of Science, \\
Kyoto University, 
Kitashirakawaoiwake, Sakyo, Kyoto 606-8502, Japan\\
        E-mail: \email{suganuma@scphys.kyoto-u.ac.jp}}

\author{Takumi Iritani \\
High Energy Accelerator Research Organization (KEK), 
Tsukuba, Ibaraki 305-0801, Japan}

\abstract{
In lattice QCD formalism, 
we derive an analytical gauge-invariant relation 
between the Polyakov loop $\langle L_P \rangle$ and 
the Dirac eigenvalues $\lambda_n$ in QCD, i.e., 
$\langle L_P \rangle \propto 
\sum_n \lambda_n^{N_t -1} \langle n|\hat U_4|n \rangle$, 
by considering ${\rm Tr} (\hat{U}_4\hat{\Slash{D}}^{N_t-1})$ 
on a temporally odd-number lattice, 
where the temporal lattice size $N_t$ is odd. 
This formula is a Dirac spectral representation of 
the Polyakov loop in terms of Dirac eigenmodes $|n\rangle$.
We here use an ordinary square lattice with 
the normal (nontwisted) periodic boundary condition 
for link-variables $U_\mu(s)$ in the temporal direction. 
From this relation, one can estimate each contribution 
of the Dirac eigenmode to the Polyakov loop.
Because of the factor $\lambda_n^{N_t -1}$ in the Dirac spectral sum, 
this analytical relation generally indicates quite small contribution of 
low-lying Dirac modes to the Polyakov loop 
in both confined and deconfined phases, 
while the low-lying Dirac modes are essential for chiral symmetry breaking. 
Also in lattice QCD calculations in confined and deconfined phases, 
we numerically confirm the analytical relation, 
non-zero finiteness of $\langle n|\hat U_4|n \rangle$ for each Dirac mode, 
and negligibly small contribution of 
low-lying Dirac modes to the Polyakov loop, i.e.,  
the Polyakov loop is almost unchanged 
even by removing low-lying Dirac-mode contribution from 
the QCD vacuum generated by lattice QCD simulations. 
Thus, we conclude that low-lying Dirac modes 
are not essential modes for confinement, 
which indicates no direct one-to-one correspondence between 
confinement and chiral symmetry breaking in QCD.
}

\FullConference{From quarks and gluons to hadronic matter: A bridge too far?\\
                 2-6 September, 2013\\
                 European Centre for Theoretical Studies in Nuclear Physics and Related Areas
(ECT*), Villazzano, Trento (Italy)}

\begin{document}

\section{Introduction: Are color confinement and CSB one-to-one in QCD?}

Quantum chromodynamics (QCD) has two outstanding nonperturbative phenomena of 
color confinement and spontaneous chiral-symmetry breaking \cite{NJL61} 
in the low-energy region, and their derivation is one of the most important 
problems in theoretical physics. 
For quark confinement, 
the Polyakov loop $\langle L_P \rangle$ is 
a typical order parameter, and 
relates to the single-quark free energy $E_q$ as 
$\langle L_P \rangle \propto e^{-E_q/T}$ 
at temperature $T$. 
$\langle L_P \rangle$ is also an order parameter 
of $Z_{N_c}$ center symmetry in QCD \cite{Rothe12}.
For chiral symmetry breaking, 
the standard order parameter 
is the chiral condensate $\langle \bar qq \rangle$, 
and low-lying Dirac modes play the essential role, 
as the Banks-Casher relation \cite{BC80} indicates.

The relation between confinement and chiral symmetry breaking 
is also one of the important physical issues 
\cite{SST95,M95W95,G06BGH07,S111213,GIS12,IS13,SDI13,DSI13}, 
and there are several circumstantial evidence on their correlation. 
For example, lattice QCD simulations have shown 
almost coincidence between deconfinement and chiral-restoration temperatures 
\cite{Rothe12,K02}, although slight difference of about 25MeV between them 
is pointed out in some recent lattice QCD studies \cite{AFKS06}.
Their correlation is also suggested 
in terms of QCD-monopoles \cite{SST95,M95W95}, 
which topologically appear in QCD in the maximally Abelian gauge 
\cite{KSW87,SNW94,STSM95,AS99,IS9900}, 
leading to the dual-superconductor picture \cite{N74tH81}. 
Actually, by removing the monopoles from the QCD vacuum 
generated in lattice QCD, 
confinement and chiral symmetry breaking are 
simultaneously lost \cite{M95W95}, 
as schematically shown in Fig.1.
This fact indicates an important role of monopoles 
to both confinement and chiral symmetry breaking, and thus 
these two phenomena seem to be related via the monopole. 
However, in spite of the essential role of monopoles, 
the direct relation of confinement and chiral symmetry breaking 
is still unclear.

\begin{figure}[ht]
\vspace{-0.14cm}
\begin{center}
\includegraphics[scale=0.3]{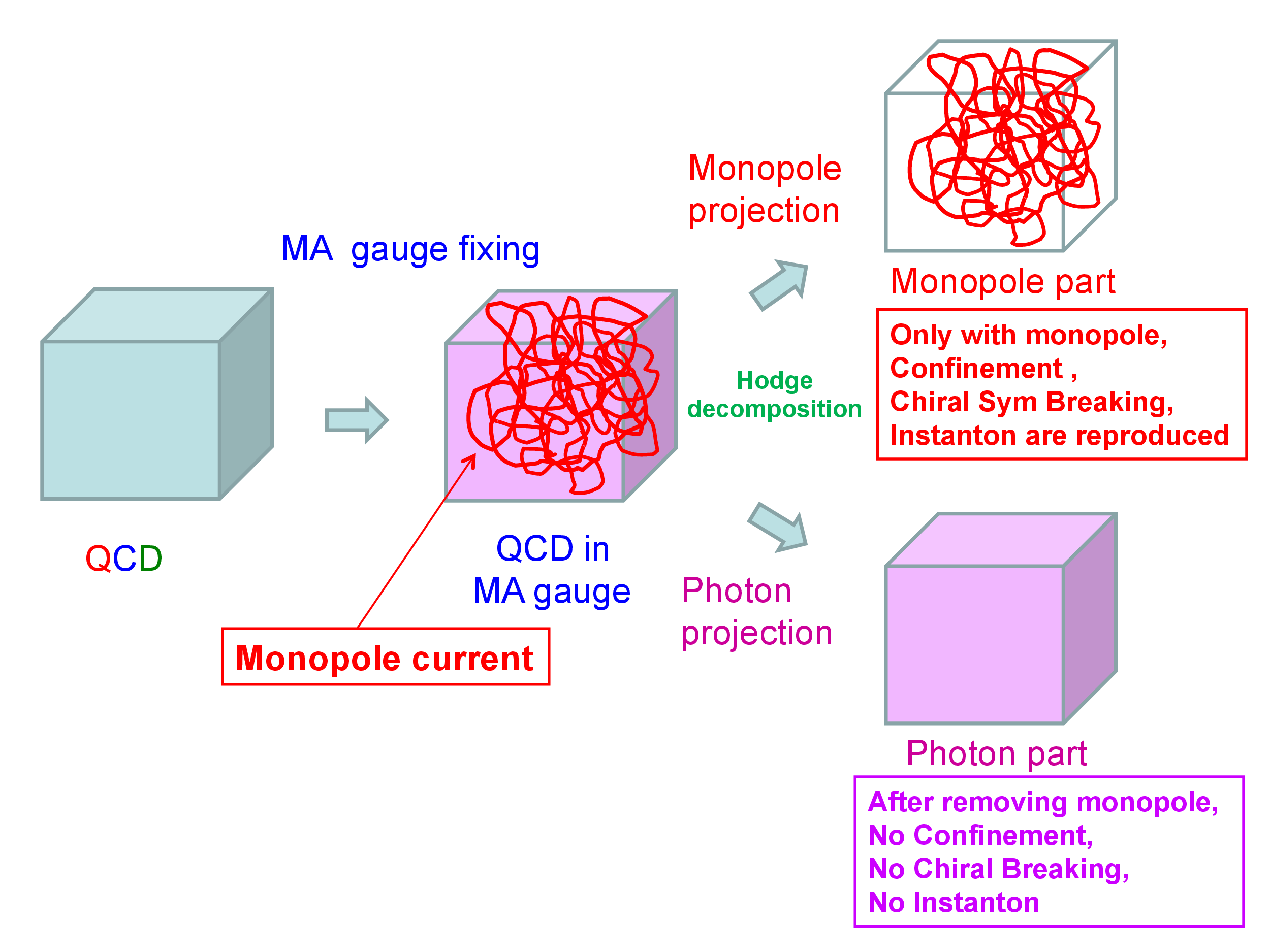}
\hspace{0.4cm} \includegraphics[scale=0.25]{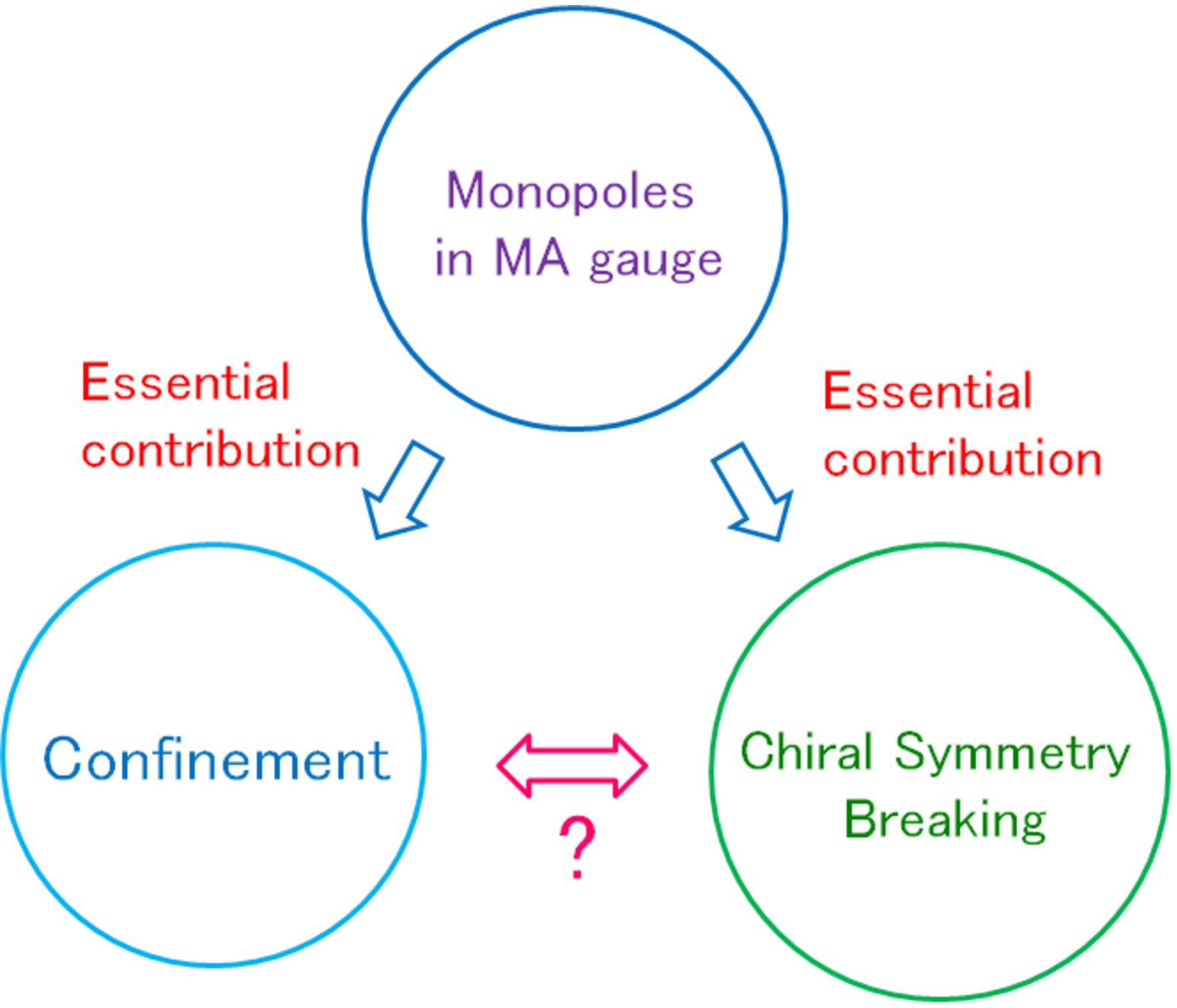}
\vspace{-0.17cm}
\caption{
The role of monopoles to nonperturbative QCD. 
In the MA gauge, QCD becomes Abelian-like due to the 
large off-diagonal gluon mass of about 1GeV \cite{AS99}, and 
monopole current topologically appears \cite{KSW87,SNW94}. 
By the Hodge decomposition, the QCD system can be divided into 
the monopole part and the photon part. 
The monopole part has confinement \cite{SNW94},
chiral symmetry breaking \cite{M95W95} and instantons \cite{STSM95},
while the photon part does not have all of them, 
as shown in lattice QCD. 
In spite of the essential role of monopoles, 
the direct relation of confinement and chiral symmetry breaking 
is still unclear.
}
\end{center}
\vspace{-0.40cm}
\end{figure}

Then, we have a question. {\it If only the relevant ingredient of 
chiral symmetry breaking is carefully removed from the QCD vacuum, 
how will be quark confinement?}

To obtain the answer, 
in this paper, we derive an analytical relation between 
the Polyakov loop and the Dirac modes in temporally odd-number lattice QCD, 
where the temporal lattice size is odd, 
and discuss the relation between confinement and chiral symmetry breaking.

\section{Lattice QCD formalism for Dirac operator, Dirac eigenvalues and Dirac modes}

First, we clarify the mathematical condition of 
lattice QCD formalism adopted in this study \cite{SDI13,DSI13}.
We use an ordinary Euclidean square lattice with spacing $a$ and 
size $V \equiv N_s^3 \times N_t$.
The normal (nontwisted) periodic boundary condition is used 
for the link-variable $U_\mu(s)={\rm e}^{iagA_\mu(s)}$ 
in the temporal direction, 
which is physically required at finite temperatures.
We take SU($N_c$) as the gauge group, 
although any gauge group $G$ can be taken 
for most arguments in this paper. 

Note that, in our studies, we just consider 
the mathematical expansion by eigenmodes $|n \rangle$ of 
the Dirac operator $\Slash D=\gamma_\mu D_\mu$, 
using the completeness of $\sum_n|n \rangle \langle n|=1$. 
In general, instead of $\Slash D$, 
one can consider any (anti)hermitian operator, e.g., $D^2=D_\mu D_\mu$, 
and the expansion in terms of its eigenmodes \cite{S111213}. 
In this paper, to consider chiral symmetry breaking, 
we adopt $\Slash D$ and the expansion by its eigenmodes.

In lattice QCD, the Dirac operator 
$\Slash D = \gamma_\mu D_\mu$ is expressed with 
$U_\mu(s)={\rm e}^{iagA_\mu(s)}$ as
\begin{eqnarray}
 \Slash{D}_{s,s'} 
 \equiv \frac{1}{2a} \sum_{\mu=1}^4 \gamma_\mu 
\left[ U_\mu(s) \delta_{s+\hat{\mu},s'}
 - U_{-\mu}(s) \delta_{s-\hat{\mu},s'} \right],
\end{eqnarray}
with $U_{-\mu}(s)\equiv U^\dagger_\mu(s-\hat \mu)$ and 
lattice unit vector $\hat\mu$.
Taking hermitian $\gamma_\mu=\gamma_\mu^\dagger$, 
the Dirac operator $\Slash D$ is anti-hermitian and satisfies 
$\Slash D_{s',s}^\dagger=-\Slash D_{s,s'}$.
We introduce the normalized Dirac eigen-state $|n \rangle$ as 
\begin{eqnarray}
\Slash D |n\rangle =i\lambda_n |n \rangle, \qquad
\langle m|n\rangle=\delta_{mn}, 
\end{eqnarray}
with the Dirac eigenvalue $i\lambda_n$ ($\lambda_n \in {\bf R}$).
Because of $\{\gamma_5,\Slash D\}=0$, the state 
$\gamma_5 |n\rangle$ is also an eigen-state of $\Slash D$ with the 
eigenvalue $-i\lambda_n$. 
Here, the Dirac eigen-state $|n \rangle$ 
satisfies the completeness of 
\begin{eqnarray}
\sum_n |n \rangle \langle n|=1.
\label{eq:DCS}
\end{eqnarray}
The Dirac eigenfunction $\psi_n(s)\equiv\langle s|n \rangle$ satisfies 
$\Slash D \psi_n(s)=i\lambda_n \psi_n(s)$, i.e.,
\begin{eqnarray}
\frac{1}{2a} \sum_{\mu=1}^4 \gamma_\mu
[U_\mu(s)\psi_n(s+\hat \mu)-U_{-\mu}(s)\psi_n(s-\hat \mu)]
=i\lambda_n \psi_n(s).
\end{eqnarray}
By the gauge transformation of 
$U_\mu(s) \rightarrow V(s) U_\mu(s) V^\dagger (s+\hat\mu)$, 
$\psi_n(s)$ is gauge-transformed as 
\begin{eqnarray}
\psi_n(s)\rightarrow V(s) \psi_n(s),
\label{eq:GTDwf}
\end{eqnarray}
which is the same as that of the quark field.
(To be strict, there can appear an irrelevant $n$-dependent 
global phase factor $e^{i\varphi_n[V]}$, 
according to arbitrariness of the phase in 
the basis $|n \rangle$ \cite{GIS12}.)

The spectral density $\rho(\lambda)$ 
of the Dirac operator $\Slash D$ relates to chiral symmetry breaking, e.g.,
the zero-eigenvalue density $\rho(0)$ leads to 
$\langle\bar qq \rangle$ (Banks-Casher's relation) \cite{BC80}.
In fact, the low-lying Dirac modes are regarded as the essential modes 
responsible to chiral symmetry breaking in QCD.

Here, we take the operator formalism 
in lattice QCD \cite{S111213,GIS12,IS13} 
by introducing the link-variable operator $\hat U_{\pm \mu}$ 
defined by the matrix element of 
\begin{eqnarray}
\langle s |\hat U_{\pm \mu}|s' \rangle 
=U_{\pm \mu}(s)\delta_{s\pm \hat \mu,s'}.
\end{eqnarray}
With the link-variable operator, 
the Dirac operator and covariant derivative 
are simply expressed as 
\begin{eqnarray}
\Slash{\hat D}
=\frac{1}{2a}\sum_{\mu=1}^{4} \gamma_\mu (\hat U_\mu-\hat U_{-\mu}),
\qquad 
\hat D_\mu=\frac{1}{2a}(\hat U_\mu-\hat U_{-\mu}).
\label{eq:Dop}
\end{eqnarray}
The Polyakov loop is also simply written as the functional trace of 
$\hat U_4^{N_t}$, i.e.,
$
\langle L_P \rangle
=\frac{1}{N_c V} \langle {\rm Tr}_c \{\hat U_4^{N_t}\}\rangle,
$
where ``${\rm Tr}_c$'' denotes the functional trace 
of ${\rm Tr}_c \equiv \sum_s {\rm tr}_c$ including 
the trace ${\rm tr}_c$ over color index.
For large volume $V$, one can expect 
$\langle O \rangle \simeq {\rm Tr}~O/{\rm Tr}~1$ 
for any operator $O$ at each gauge configuration.
The Dirac-mode matrix element of the link-variable operator 
$\hat U_{\mu}$ can be expressed with $\psi_n(s)$ as 
\begin{eqnarray}
\langle m|\hat U_{\mu}|n \rangle=\sum_s\langle m|s \rangle 
\langle s|\hat U_{\mu}|s+\hat \mu \rangle \langle s+\hat \mu|n\rangle
=\sum_s \psi_m^\dagger(s) U_\mu(s)\psi_n(s+\hat \mu),
\end{eqnarray}
which is gauge invariant, 
because of (\ref{eq:GTDwf}), 
apart from an irrelevant global phase factor \cite{GIS12}.

\section{Previous numerical study: 
Dirac-mode expansion and Dirac-mode projection} 

We here review our previous studies \cite{S111213,GIS12,IS13} 
on ``Dirac-mode expansion'', ``Dirac-mode projection'' 
where the Dirac-mode space is restricted, 
and the role of low-lying Dirac modes 
to confinement in SU(3) lattice QCD. 

From the completeness of the Dirac-mode basis, 
$\sum_n|n\rangle \langle n|=1$, 
arbitrary operator $\hat O$ can be expanded 
in terms of the Dirac-mode basis $|n \rangle$ as 
$
\hat O=\sum_n \sum_m |n \rangle \langle n|\hat O|m \rangle \langle m|.
$
With this relation, we consider the Dirac-mode expansion and 
Dirac-mode projection. 
We define the projection operator 
$
\hat P\equiv \sum_{n \in A}|n\rangle \langle n|,
$
which restricts the Dirac-mode space to 
its arbitrary subset $A$. 
Using the projection operator $\hat P$, we define 
the Dirac-mode projected link-variable operator 
$
\hat U^P_\mu \equiv \hat P \hat U_\mu \hat P
=\sum_{m \in A}\sum_{n \in A} 
|m\rangle \langle m|\hat U_\mu|n\rangle \langle n|,
$
the Dirac-mode projected Wilson-loop operator 
$\hat W^P \equiv \prod_{k=1}^N \hat U^P_{\mu_k}$, 
the Dirac-mode projected inter-quark potential 
$
V^P(R)\equiv -\lim_{T \to \infty} \frac{1}{T}
{\rm ln} \langle {\rm Tr}_c \ \hat W^P(R,T)\rangle
$
from the $R \times T$ rectangular Wilson loop, 
and the Dirac-mode projected Polyakov loop  
$
\langle L_{P}\rangle_{\rm proj.} \equiv
\frac{1}{N_c V} \langle {\rm Tr}_c \ (\hat{U}_4^P)^{N_t} \rangle.
$

In Refs.\cite{S111213,GIS12}, 
we use SU(3) quenched lattice QCD at $\beta=5.6$ 
(i.e., $a\simeq 0.25{\rm fm}$) on $6^4$, 
and take the IR-cutoff $\Lambda_{\rm IR}=0.5a^{-1}\simeq 0.4{\rm GeV}$ 
for the Dirac modes, which leads an extreme reduction of 
the chiral condensate as 
$\langle \bar qq\rangle_{\Lambda_{\rm IR}}/
\langle \bar qq\rangle \simeq 0.02$
for the current quark mass $m \simeq 5{\rm MeV}$.
Figure~2 shows the IR-Dirac-mode-cut 
Wilson loop $\langle {\rm Tr}_c \hat W^P(R,T)\rangle$, 
the IR-cut inter-quark potential $V^P(R)$, and 
the IR-Dirac-mode-cut Polyakov loop $\langle L_P \rangle_{\rm IR\hbox{-}cut}$, 
after the removal of the low-lying Dirac modes.

\begin{figure}[h]
\begin{center}
\includegraphics[scale=0.4]{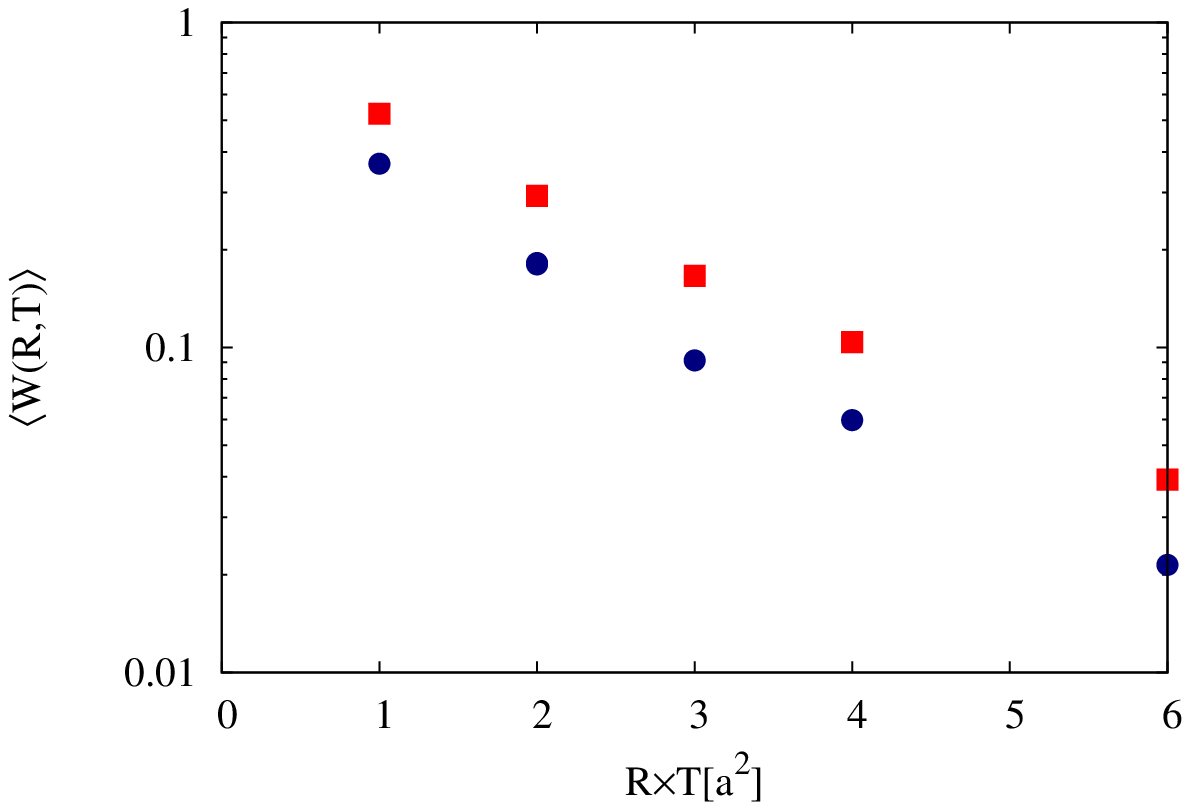}
\includegraphics[scale=0.4]{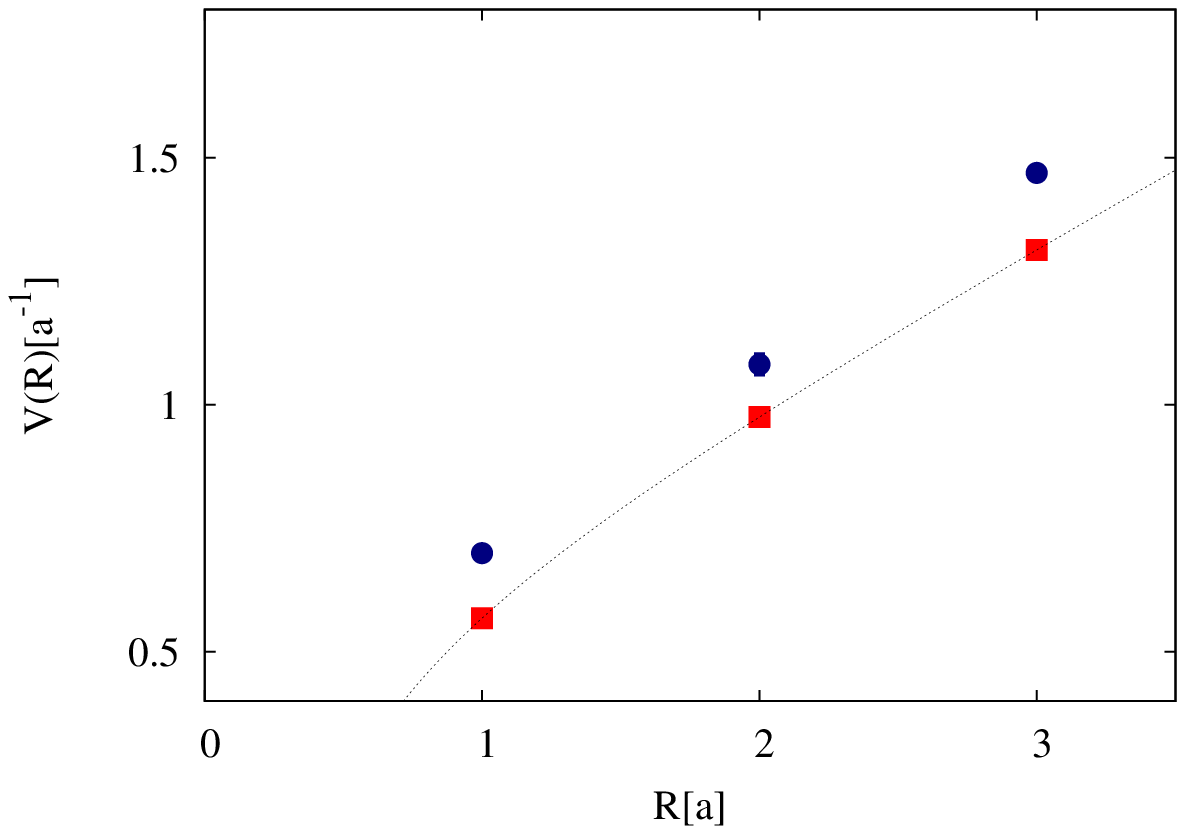}
\includegraphics[scale=0.4]{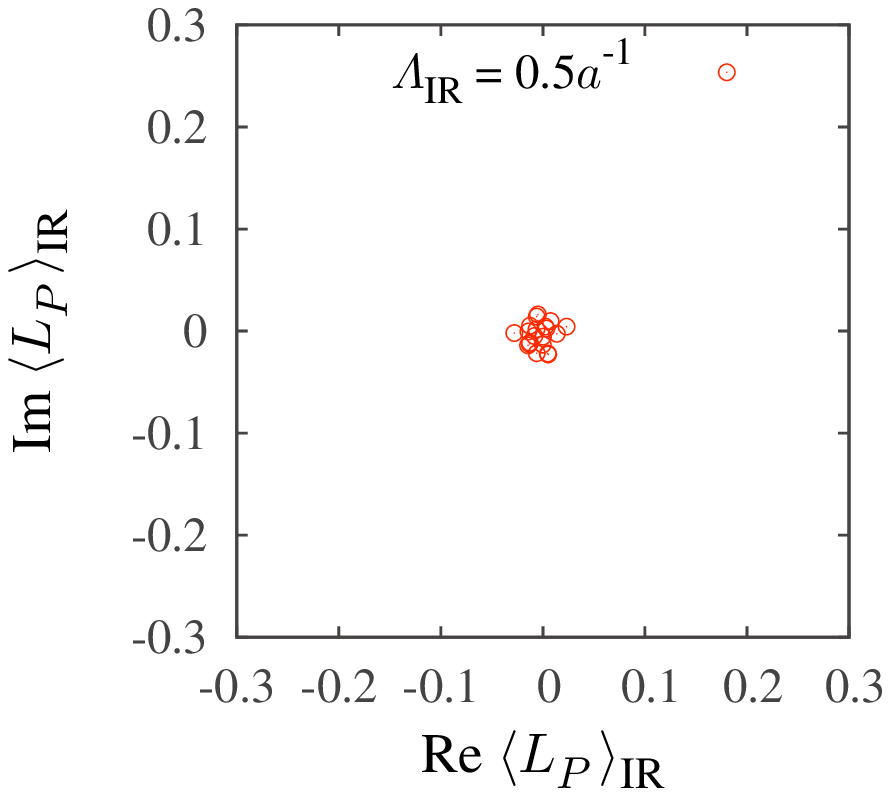}
\caption{
Lattice QCD results \cite{S111213,GIS12}
after the removal of low-lying Dirac modes  
below the IR-cutoff $\Lambda_{\rm IR}=0.5a^{-1} \simeq 0.4{\rm GeV}$.
(a) The IR-cut Wilson loop $\langle {\rm Tr}_c~W^P(R,T) \rangle$ (circle) 
plotted against $R \times T$. 
The slope $\sigma^P$ is almost the same as that of 
the original Wilson loop (square).
(b) The IR-cut inter-quark potential $V^P(R)$ (circle). 
$V^P(R)$ is almost unchanged from the original one (square),
apart from an irrelevant constant.
(c) The scatter plot of the IR-cut Polyakov loop 
$\langle L_P \rangle_{\rm IR\hbox{-}cut}$.
$\langle L_P \rangle_{\rm IR\hbox{-}cut}\simeq 0$ means $Z_3$-unbroken 
confinement phase.
}
\end{center}
\vspace{-0.5cm}
\end{figure}

Remarkably, 
even after removing the coupling to the low-lying Dirac modes, 
the IR-Dirac-mode-cut  Wilson loop obeys the area law as 
$\langle W^P(R,T)\rangle \propto e^{-\sigma^P RT}$, 
and the slope $\sigma^P$, i.e., the string tension, 
is almost unchanged as $\sigma^P \simeq \sigma$.
As shown in Fig.2(b), 
the IR-cut inter-quark potential $V^P(R)$ 
is almost unchanged from the original one, 
apart from an irrelevant constant. 
Also from Fig.2(c), we find that the IR-Dirac-mode-cut 
Polyakov loop is almost zero, $\langle L_P \rangle_{\rm IR\hbox{-}cut} \simeq 0$, 
which means $Z_3$-unbroken confinement phase.
In this way, confinement is kept 
even in the absence of low-lying Dirac modes or 
the essence of chiral symmetry breaking \cite{S111213,GIS12,IS13}. 

We also investigate the UV-cut of Dirac modes in lattice QCD, 
and find that the confining force is almost unchanged 
by the UV-cut \cite{S111213,GIS12,IS13}.
Furthermore, we examine ``intermediate(IM)-cut'' of Dirac modes, 
and obtain almost the same confining force \cite{S111213,GIS12}.

From these lattice QCD results, there is no specific region of 
the Dirac modes responsible to confinement. 
In other words, we conjecture that the ``seed'' of confinement is 
distributed not only in low-lying Dirac modes but also 
in a wider region of the Dirac-mode space. 

\section{Analytical relation of the Polyakov loop and Dirac modes
in odd-$N_t$ lattice QCD}

Now, we consider temporally odd-number lattice QCD, 
where the temporal lattice size $N_t$ is odd \cite{SDI13,DSI13}.
Here, we use an ordinary square lattice 
with the normal (nontwisted) periodic boundary condition 
for link-variables $U_\mu(s)$ in the temporal direction. 
The spatial lattice size $N_s$ is taken to be larger than $N_t$, 
i.e., $N_s > N_t$. 
Note that, in the continuum limit of $a \rightarrow 0$ and 
$N_t \rightarrow \infty$, 
any large number $N_t$ gives the same physical result.
Then, to use the odd-number lattice is no problem.

\begin{figure}[h]
\begin{center}
\includegraphics[scale=0.3]{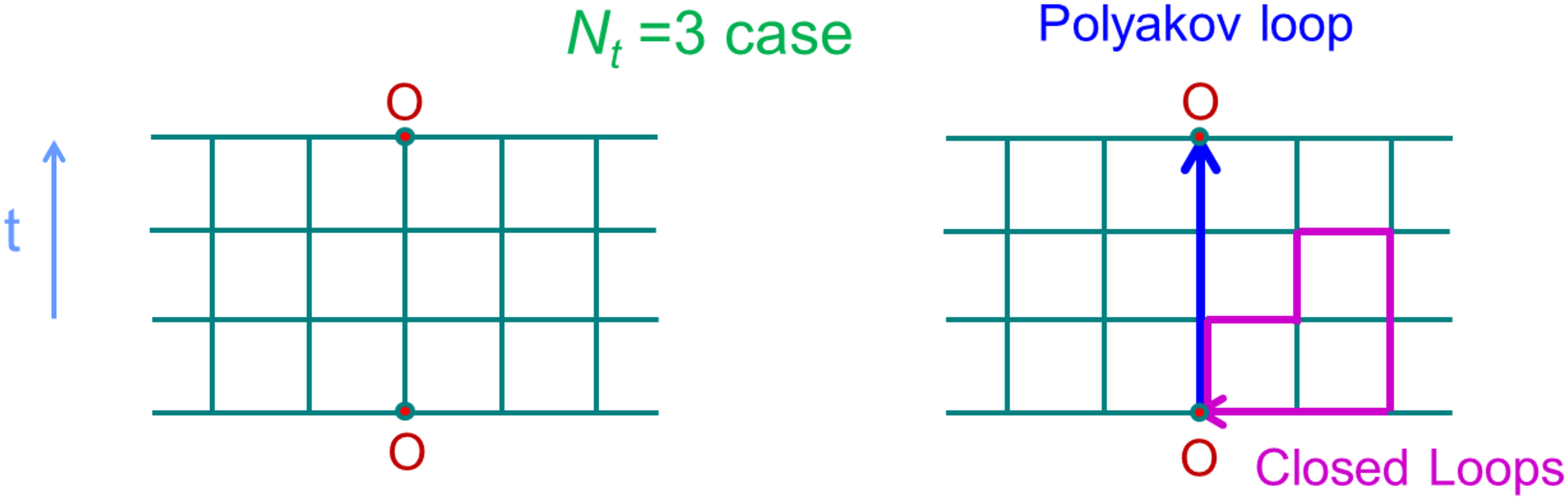}
\caption{
An example of the temporally odd-number lattice ($N_t=3$ case).
Only gauge-invariant quantities such as 
closed loops and the Polyakov loop survive in QCD, 
after taking the expectation value, i.e., the gauge-configuration average.
Geometrically, closed loops have even-number links on the square lattice.
}
\end{center}
\vspace{-0.3cm}
\end{figure}

As a general mathematical argument 
of the Elitzur theorem \cite{Rothe12}, 
only gauge-invariant quantities 
such as closed loops and the Polyakov loop survive in QCD.
In fact, all the non-closed lines are gauge-variant 
and their expectation values are zero.
Note here that any closed loop  
needs even-number link-variables on the square lattice,
except for the Polyakov loop \cite{SDI13}. (See Fig.3.)

In temporally odd-number lattice QCD \cite{SDI13,DSI13}, 
we consider the functional trace of 
\begin{eqnarray}
I\equiv {\rm Tr}_{c,\gamma} (\hat{U}_4\hat{\Slash{D}}^{N_t-1}), 
\label{eq:FT}
\end{eqnarray}
where 
${\rm Tr}_{c,\gamma}\equiv \sum_s {\rm tr}_c 
{\rm tr}_\gamma$ includes 
${\rm tr}_c$ 
and the trace ${\rm tr}_\gamma$ over spinor index.
Its expectation value 
\begin{eqnarray}
 \langle I\rangle=\langle {\rm Tr}_{c,\gamma} (\hat{U}_4\hat{\Slash{D}}^{N_t-1})\rangle 
\label{eq:FTV}
\end{eqnarray}
is obtained as the gauge-configuration average in lattice QCD.
When the volume $V$ is enough large, one can expect 
$\langle O \rangle \simeq {\rm Tr}~O/{\rm Tr}~1$ 
for any operator $O$ even in each gauge configuration.

From Eq.(\ref{eq:Dop}), 
$\hat U_4\Slash{\hat D}^{N_t-1}$ 
can be expressed as a sum of products of $N_t$ link-variable operators, 
since the Dirac operator $\Slash{\hat D}$ 
includes one link-variable operator in each direction of $\pm \mu$.
In fact, $\hat U_4\Slash{\hat D}^{N_t-1}$ 
includes ``many trajectories'' with the total length $N_t$ 
(in lattice unit) on the square lattice, as shown in Fig.4.
Note that 
all the trajectories with the odd-number length $N_t$ 
cannot form a closed loop 
on the square lattice, and give gauge-variant contribution, 
except for the Polyakov loop \cite{SDI13}.

\begin{figure}[h]
\begin{center}
\includegraphics[scale=0.3]{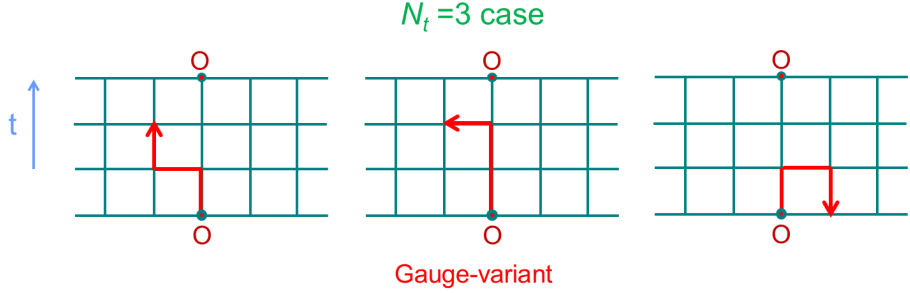}
\vspace{-0.1cm}
\caption{
Partial examples of the trajectories stemming from 
$\langle {\rm Tr}_{c,\gamma}(\hat U_4\Slash{\hat D}^{N_t-1})\rangle$. 
For each trajectory, the total length is $N_t$, and 
the ``first step'' is positive temporal direction 
corresponding to $\hat U_4$.
All the trajectories with the odd-number length $N_t$ 
cannot form a closed loop on the square lattice,
so that they are gauge-variant and give no contribution 
in $\langle {\rm Tr}_{c,\gamma}(\hat U_4 \Slash{\hat D}^{N_t-1})\rangle$, 
except for the Polyakov loop.
}
\end{center}
\vspace{-0.2cm}
\end{figure}

Hence, among the trajectories stemming from 
$\langle {\rm Tr}_{c,\gamma}(\hat U_4\Slash{\hat D}^{N_t-1}) \rangle$, 
all the non-loop trajectories are gauge-variant and give no contribution, 
according to the Elitzur theorem \cite{Rothe12}.
Only the exception is the Polyakov loop. (Compare Figs.4 and 5.)
Note that $\langle {\rm Tr}_{c,\gamma}
(\hat U_4 \Slash{\hat D}^{N_t-1})\rangle$ 
does not include the anti-Polyakov loop $\langle L_P^\dagger \rangle$, 
since the ``first step'' is positive temporal direction 
corresponding to $\hat U_4$.

Thus, in the functional trace 
$\langle I \rangle
=\langle{\rm Tr}_{c,\gamma}(\hat U_4\Slash{\hat D}^{N_t-1})\rangle$, 
only the Polyakov-loop ingredient can survive 
as the gauge-invariant quantity, and 
$\langle I \rangle$ is proportional to the Polyakov loop $\langle L_P \rangle$.

\begin{figure}[h]
\vspace{-0.3cm}
\begin{center}
\includegraphics[scale=0.3]{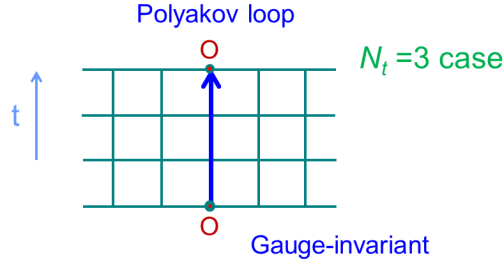}
\vspace{-0.1cm}
\caption{
Among the trajectories stemming from 
$\langle {\rm Tr}_{c,\gamma}(\hat U_4\Slash{\hat D}^{N_t-1}) \rangle$, 
only the Polyakov-loop ingredient can survive 
as the gauge-invariant quantity. 
Owing to the first factor $\hat U_4$, 
$\langle {\rm Tr}_{c,\gamma}(\hat U_4\Slash{\hat D}^{N_t-1}) \rangle$ 
does not include $\langle L_P^\dagger \rangle$.
}
\end{center}
\vspace{-0.5cm}
\end{figure}

Actually, we can mathematically derive the following relation 
\cite{SDI13}:
\begin{eqnarray}
\langle I\rangle
&=&\langle {\rm Tr}_{c,\gamma} (\hat U_4 \hat{\Slash D}^{N_t-1}) \rangle
\nonumber \\
&=&\langle {\rm Tr}_{c,\gamma} \{\hat U_4 (\gamma_4 \hat D_4)^{N_t-1}\} \rangle
\quad \qquad \qquad ~~~~~{\rm 
(
\raisebox{1.2ex}{.}\raisebox{.2ex}{.}\raisebox{1.2ex}{.} 
~only~gauge\hbox{-}invariant~terms~survive)} 
\nonumber \\
&=&4\langle {\rm Tr}_{c} (\hat U_4 \hat D_4^{N_t-1}) \rangle
\quad \qquad \qquad \qquad ~~~~~~(
\raisebox{1.2ex}{.}\raisebox{.2ex}{.}\raisebox{1.2ex}{.} 
~\gamma_4^{N_t-1}={1}, 
~{\rm tr}_\gamma {1}=4) 
\nonumber \\
&=&\frac{4}{(2a)^{N_t-1}}
\langle {\rm Tr}_{c} \{\hat U_4 (\hat U_4-\hat U_{-4})^{N_t-1}\} \rangle
\quad ~~(
\raisebox{1.2ex}{.}\raisebox{.2ex}{.}\raisebox{1.2ex}{.} 
~\hat D_4=\frac{1}{2a}(\hat U_4-\hat U_{-4}))
\nonumber \\
&=&\frac{4}{(2a)^{N_t-1}} \langle {\rm Tr}_{c} \{ \hat U_4^{N_t} \}\rangle
=\frac{12V}{(2a)^{N_t-1}}\langle L_P \rangle.
~{\rm 
(\raisebox{1.2ex}{.}\raisebox{.2ex}{.}\raisebox{1.2ex}{.} 
~only~gauge\hbox{-}invariant~terms~survive)} 
\label{eq:FTdetail}
\end{eqnarray}
We thus obtain the relation between 
$\langle I\rangle = \langle {\rm Tr}_{c,\gamma}
 (\hat U_4 \hat{\Slash D}^{N_t-1}) \rangle$ 
and the Polyakov loop $\langle L_P \rangle$,
\begin{eqnarray}
\langle I\rangle
=\langle {\rm Tr}_{c,\gamma} (\hat U_4 \hat{\Slash D}^{N_t-1}) \rangle
=\frac{12V}{(2a)^{N_t-1}}\langle L_P \rangle. 
\label{eq:FTtoPL}
\end{eqnarray}

On the other hand, we calculate the functional trace 
in Eq.(\ref{eq:FTV}) using the complete set of 
the Dirac-mode basis $|n\rangle$ satisfying $\sum_n |n\rangle \langle n|=1$, 
and find the Dirac-mode representation of 
\begin{eqnarray}
 \langle I\rangle=\sum_n\langle n|\hat{U}_4\Slash{\hat{D}}^{N_t-1}|n\rangle
=i^{N_t-1}\sum_n\lambda_n^{N_t-1}\langle n|\hat{U}_4| n \rangle. 
\label{eq:FTtoD}
\end{eqnarray}
Combing Eqs.(\ref{eq:FTtoPL}) and (\ref{eq:FTtoD}), we obtain the analytical 
relation between the Polyakov loop $ \langle L_P \rangle$ 
and the Dirac eigenvalues $i\lambda_n$ \cite{SDI13} in QCD: 
\begin{eqnarray}
\langle L_P \rangle=\frac{(2ai)^{N_t-1}}{12V}
\sum_n\lambda_n^{N_t-1}\langle n|\hat{U}_4| n \rangle. 
\label{eq:PLvsD}
\end{eqnarray}
This is a direct relation between the Polyakov loop $\langle L_P\rangle$ 
and the Dirac modes in QCD, i.e., 
a ``Dirac spectral representation of the Polyakov loop'', 
and is mathematically valid on the temporally odd-number lattice 
in both confined and deconfined phases. 
Based on Eq.(\ref{eq:PLvsD}), 
we can investigate each Dirac-mode contribution 
to the Polyakov loop individually, e.g., 
by evaluating each contribution specified by $n$ 
numerically in lattice QCD.
In particular, by paying attention to low-lying Dirac modes 
in Eq.(\ref{eq:PLvsD}), the relation between confinement 
and chiral symmetry breaking can be discussed in QCD.

As a remarkable fact, because of the factor $\lambda_n^{N_t -1}$, 
the contribution from 
low-lying Dirac-modes with $|\lambda_n|\simeq 0$ 
is negligibly small in the Dirac spectral sum of RHS in Eq.(\ref{eq:PLvsD}),
compared to the other Dirac-mode contribution. 
In fact, the low-lying Dirac modes have little contribution 
to the Polyakov loop, regardless of confined or deconfined phase 
\cite{SDI13,DSI13}.
(This result agrees with the previous numerical 
lattice results that confinement properties are almost unchanged by 
removing low-lying Dirac modes from the QCD vacuum \cite{S111213,GIS12,IS13}.)
Thus, we conclude from the relation (\ref{eq:PLvsD}) 
that low-lying Dirac modes are not essential modes for confinement, 
which indicates no direct one-to-one correspondence between 
confinement and chiral symmetry breaking in QCD.

Here, we give several comments on the relation (\ref{eq:PLvsD}) in order.
\begin{enumerate}
\item 
Equation (\ref{eq:PLvsD}) is a manifestly gauge-invariant relation. 
Actually, the matrix element $\langle n |\hat U_4|n\rangle$ 
can be expressed with 
the Dirac eigenfunction $\psi_n(s)$ and 
the temporal link-variable $U_4(s)$ as 
\begin{eqnarray}
\langle n |\hat U_4|n\rangle =
\sum_s \langle n |s \rangle \langle s 
|\hat U_4| s+\hat t \rangle \langle s+\hat t|n\rangle
=\sum_s \psi_n^\dagger (s)U_4(s) \psi_n(s+\hat t),
\end{eqnarray}
and each term $\psi_n^\dagger (s)U_4(s) \psi_n(s+\hat t)$ 
is manifestly gauge invariant, because of 
the gauge transformation property (\ref{eq:GTDwf}).
[Global phase factors also cancel exactly 
between $\langle n|$ and $|n \rangle$.] 
\item
In RHS of Eq.(\ref{eq:PLvsD}), 
there is no cancellation between chiral-pair Dirac eigen-states, 
$|n \rangle$ and $\gamma_5|n \rangle$, because $(N_t-1)$ is even, i.e., 
$(-\lambda_n)^{N_t-1}=\lambda_n^{N_t-1}$, and  
$\langle n |\gamma_5 \hat U_4 \gamma_5|n\rangle
=\langle n |\hat U_4|n\rangle$. 
\vspace{0.2cm}
\item
Even in the presence of a possible 
multiplicative renormalization factor for the Polyakov loop,
the contribution from the low-lying Dirac modes (or 
the small $|\lambda_n|$ region) is relatively negligible, 
compared to other Dirac-mode contribution 
in the sum of RHS in Eq.(\ref{eq:PLvsD}). 
\item
For the arbitrary color number $N_c$, 
Eq.(\ref{eq:PLvsD}) is true and applicable in the SU($N_c$) gauge theory.
\item
If RHS in Eq.(\ref{eq:PLvsD}) {\it were} not a sum but a product, 
low-lying Dirac modes should have given an important contribution 
to the Polyakov loop as a crucial reduction factor of $\lambda_n^{N_t-1}$. 
In the sum, however, the contribution ($\propto \lambda_n^{N_t-1}$) 
from the small $|\lambda_n|$ region 
is negligible. 
\item
Even if $\langle n |\hat U_4|n\rangle$ behaves as the $\delta$-function 
$\delta(\lambda)$, the factor $\lambda_n^{N_t-1}$ is still crucial 
in RHS of Eq.(\ref{eq:PLvsD}), 
because of $\lambda \delta(\lambda)=0$. 
\item
The relation (\ref{eq:PLvsD}) is correct regardless of 
presence or absence of dynamical quarks, 
although dynamical quark effects appear in $\langle L_P\rangle$, 
the Dirac eigenvalue distribution $\rho(\lambda)$ and 
$\langle n |\hat U_4|n\rangle$.
\item
The relation (\ref{eq:PLvsD}) is correct also 
at finite baryon density and finite temperature.
\item
Equation (\ref{eq:PLvsD}) obtained on the odd-number lattice 
is correct in the continuum limit of $a \rightarrow 0$ 
and $N_t \rightarrow \infty$, since 
any number of large $N_t$ gives the same physical result.
\end{enumerate}

Note that most of the above arguments 
can be numerically confirmed by lattice QCD calculations. 
Using actual lattice QCD calculations at the quenched level, 
we numerically confirm the analytical relation (\ref{eq:PLvsD}), 
non-zero finiteness of $\langle n|\hat U_4|n\rangle$ for each Dirac mode, 
and the negligibly small contribution of 
low-lying Dirac modes to the Polyakov loop, 
in both confined and deconfined phases \cite{SDI13,DSI13}, 
as will be shown in Sec.5. 
Although we numerically find an interesting drastic change of 
the behavior of $\langle n|\hat U_4|n\rangle$ 
between confined and deconfined phases, 
we find also quite small contribution of 
low-lying Dirac modes to the Polyakov loop.

\section{Modified KS formalism for temporally odd-number lattice}

The Dirac operator $\Slash{D}$ has a large dimension of 
$(4 \times N_{\rm c}\times V)^2$, and hence 
the numerical cost for solving the Dirac eigenvalue equation is quite huge.
This numerical cost can be partially reduced 
using the Kogut-Susskind (KS) formalism \cite{Rothe12,GIS12,DSI13,KS75}.
However, the original KS formalism can be applied only to the ``even lattice'' 
where all the lattice sizes $N_\mu$ are even number. 
In this section, we modify the KS formalism 
to be applicable to the odd-number lattice \cite{DSI13}. 
Using the ``modified KS formalism'', we can reduce the numerical cost 
in the case of the temporally odd-number lattice.

In the original KS formalism for even lattices, 
using the matrix 
$
 T(s)\equiv\gamma_1^{s_1}\gamma_2^{s_2}\gamma_3^{s_3}\gamma_4^{s_4}, 
$
all the $\gamma$-matrices can be diagonalized as 
$
T^\dagger(s)\gamma_\mu T(s\pm\hat{\mu})=\eta_\mu(s){\bf 1}, 
$
where $\eta_\mu(s)$ is the staggered phase,
$
 \eta_1(s)\equiv 1, \ \ \eta_\mu(s)\equiv (-1)^{s_1+\cdots+s_{\mu-1}} 
\ (\mu \geq 2). 
$
Then, the Dirac operator $\Slash{D}$ is spin-diagonalized as 
\begin{eqnarray}
 \sum_\mu T^\dagger(s) \gamma_\mu D_\mu T(s+ \hat \mu) = {\rm diag}(\eta_\mu D_\mu,\eta_\mu D_\mu,\eta_\mu D_\mu,\eta_\mu D_\mu), 
\label{Eq:TDiracT}
\end{eqnarray}
where $\eta_\mu D_\mu$ is the KS Dirac operator given by
\begin{eqnarray}
(\eta_\mu D_\mu)_{ss'}=\frac{1}{2a}\sum_{\mu=1}^{4}\eta_\mu(s)\left[U_\mu(s)\delta_{s+\hat{\mu},s'}-U_{-\mu}(s)\delta_{s-\hat{\mu},s'}\right]. \label{Eq:KSDiracOp}
\end{eqnarray}
Equation (\ref{Eq:TDiracT}) shows fourfold degeneracy of the Dirac eigenvalue 
relating to the spiror structure, and 
then all the eigenvalues $i\lambda_n$ are obtained by solving 
the reduced Dirac eigenvalue equation
\begin{eqnarray}
\eta_\mu D_\mu|n) =i\lambda_n|n ). \label{Eq:KSEigenEqOp}
\end{eqnarray}
Using the eigenfunction $\chi_n(s)\equiv\langle s|n )$ 
of the KS Dirac operator, 
the explicit form of Eq.(\ref{Eq:KSEigenEqOp}) reads
\begin{eqnarray}
\frac{1}{2a}\sum_{\mu=1}^4 
\eta_\mu(x)[U_\mu(x) \chi_n(x+\hat \mu)-U_{-\mu}(x)
\chi_n(x-\hat \mu)] =i\lambda_n\chi_n(x), \label{Eq:KSEigenEq}
\end{eqnarray}
where the relation between the Dirac eigenfunction 
$\psi_n(s)$ and the spinless eigenfunction $\chi_n(s)$ is
\begin{eqnarray}
\psi_n(s)=T(s)\chi_n(s). \label{Eq:PsiChiEven}
\end{eqnarray}

Note here that the original KS formalism is applicable only to even lattices 
in the presence of the periodic boundary condition \cite{DSI13}. 
In fact, the periodic boundary condition requires 
\begin{eqnarray}
T(s+N_\mu\hat{\mu})=T(s) \ (\mu=1,2,3,4),
\end{eqnarray}
however, it is satisfied only on even lattices.
Note also that, while the spatial boundary condition can be changed arbitrary, 
the temporal periodic boundary condition physically appears and 
cannot be changed at finite temperatures. 
Thus, the original KS formalism cannot be applied 
on the temporally odd-number lattice.

Now, we consider the temporally odd-number lattice, 
with all the spatial lattice size being even. 
Instead of the matrix $T(s)$, we introduce a new matrix \cite{DSI13}
\begin{eqnarray}
 M(s)\equiv\gamma_1^{s_1}\gamma_2^{s_2}\gamma_3^{s_3}\gamma_4^{s_1+s_2+s_3}, 
\label{Eq:M}
\end{eqnarray}
where the exponent of $\gamma_4$ differs from $T(s)$.
As a remarkable feature, the requirement from 
the periodic boundary condition is satisfied 
on the temporally odd-number lattice \cite{DSI13}:
\begin{eqnarray}
M(s+N_\mu\hat{\mu})=M(s) \ (\mu=1,2,3,4).
\end{eqnarray}
Using the matrix $M(s)$, all the $\gamma$-matrices 
are transformed to be proportional to $\gamma_4$:
\begin{eqnarray}
 M^\dagger(s)\gamma_\mu M(s\pm\hat{\mu})=\eta_\mu(s)\gamma_4, 
\label{Eq:MgammaM}
\end{eqnarray}
where $\eta_\mu(x)$ is the staggered phase.
In the Dirac representation, $\gamma_4$ is diagonal as 
\begin{eqnarray}
 \gamma_4={\rm diag}(1,1,-1,-1) \ \ \ (\rm{Dirac \ representation}),
\label{Eq:gamma_4}
\end{eqnarray}
and we take the Dirac representation.
Thus, we can spin-diagonalize the Dirac operator $\Slash{D}$ in the case of 
the temporally odd-number lattice \cite{DSI13}:
\begin{eqnarray}
 \sum_\mu M^\dagger(s) \gamma_\mu D_\mu M(s+ \hat \mu) = {\rm diag}(\eta_\mu D_\mu,\eta_\mu D_\mu,-\eta_\mu D_\mu,-\eta_\mu D_\mu), \label{Eq:MDiracM}
\end{eqnarray}
where $\eta_\mu D_\mu$ is 
the KS Dirac operator given by Eq.(\ref{Eq:KSDiracOp}).
Then, for each $\lambda_n$, 
two positive modes and two negative modes appear 
relating to the spinor structure on the temporally odd-number lattice.
(Note also that the chiral partner $\gamma_5 |n \rangle$ 
gives an eigenmode with the eigenvalue $-i\lambda_n$.)
In any case, all the eigenvalues $i\lambda_n$ can be obtained by 
solving the reduced Dirac eigenvalue equation 
\begin{eqnarray}
\eta_\mu D_\mu|n) =\pm i\lambda_n|n ) \label{Eq:KSEigenEqOpOdd}
\end{eqnarray}
just like the case of even lattices.
The relation between the Dirac eigenfunction $\psi_n(s)$ and 
the sponless eigenfunction $\chi_n(s)\equiv\langle s|n ) $ 
is given by 
\begin{eqnarray}
\psi_n(s)=M(s)\chi_n(s) \label{Eq:PsiChiOdd}
\end{eqnarray}
on the temporally odd-number lattice.

\section{Numerical confirmation for the relation between Polyakov loop 
and Dirac modes}

Using the modified KS formalism, Eq.(\ref{eq:PLvsD}) is rewritten as
\begin{eqnarray}
\langle L_P \rangle =\frac{(2ai)^{N_t-1}}{3V}
\sum_n\lambda_n^{N_t-1}( n|\hat{U}_4| n ). \label{Eq:RelKS}
\end{eqnarray}
Note that the (modified) KS formalism is an exact method for diagonalizing 
the Dirac operator and is not an approximation, so that 
Eqs.(\ref{eq:PLvsD}) and (\ref{Eq:RelKS}) are completely equivalent.
In fact, the relation (\ref{eq:PLvsD}) can be confirmed 
by the numerical test of the relation (\ref{Eq:RelKS}). 

We numerically calculate LHS and RHS of the relation (\ref{Eq:RelKS}), 
respectively, and compare them \cite{DSI13}. 
We perform SU(3) lattice QCD Monte Carlo simulations with 
the standard plaquette action at the quenched level 
in both cases of confined and deconfined phases.
For the confined phase, we use $10^3\times 5$ lattice with 
$\beta\equiv2N_{\rm c}/g^2=5.6$ (i.e., $a\simeq 0.25~{\rm fm}$), 
corresponding to $T\equiv1/(N_ta)\simeq160~{\rm MeV}$.
For the deconfined phase, we use $10^3\times 3$ lattice with $\beta=5.7$ 
(i.e., $a\simeq 0.20~{\rm fm}$), corresponding to 
$T\equiv1/(N_ta)\simeq 330~{\rm MeV}$.

As the numerical result, 
comparing LHS and RHS of the relation (\ref{Eq:RelKS}), 
we find that the relation (\ref{Eq:RelKS}) is almost exact even for each 
gauge configuration in both confined and deconfined phases \cite{DSI13}.
Therefore, the relation (\ref{Eq:RelKS}) 
is satisfied also for the gauge-configuration average. 

Next, we numerically confirm that the low-lying Dirac modes 
have negligible contribution to the Polyakov loop using Eq.(\ref{Eq:RelKS}).
By checking all the Dirac modes, we find that the matrix element 
$(n|\hat U_4|n)$ is generally nonzero 
for each Dirac mode \cite{SDI13,DSI13}.
In fact, for low-lying Dirac modes, the factor $\lambda_n^{N_t-1}$ 
plays a crucial role in RHS of Eq.(\ref{Eq:RelKS}).
Since RHS of Eq.(\ref{Eq:RelKS}) is expressed as a sum of 
the Dirac-mode contribution, 
we calculate the Polyakov loop without low-lying Dirac-mode contribution as 
\begin{eqnarray}
 \langle L_P \rangle_{\rm IR\hbox{-}cut} \equiv \frac{(2ai)^{N_t-1}}{3V}
\sum_{|\lambda_n|>\Lambda_{\rm IR}}\lambda_n^{N_t-1}( n|\hat{U}_4| n ) 
\label{Eq:IRcut},
\end{eqnarray}
with the IR cut $\Lambda_{\rm IR}$ for the Dirac eigenvalue.
The chiral condensate without the contribution from the low-lying Dirac-mode  
below IR cut $\Lambda_{\rm IR}$ is given by \cite{GIS12,DSI13} 
\vspace{-0.05cm}
\begin{eqnarray}
 \langle \bar qq\rangle_{\Lambda_{\rm IR}} = -\frac{1}{V}\sum_{\lambda_n > \Lambda_{\rm IR}} \frac{2m}{\lambda_n^2+m^2}, \label{Eq:qbarqIR}
\end{eqnarray}
\vspace{-0.1cm}
where $m$ is the current quark mass.
Here, we take the IR cut of $\Lambda_{\rm IR}\simeq0.4 {\rm GeV}$.
In the confined phase, this IR Dirac-mode cut leads to 
$
{\langle \bar qq\rangle_{\Lambda_{\rm IR}}}/
{\langle \bar qq\rangle} \simeq 0.02
$
and almost chiral-symmetry restoration 
for the current quark mass $m\simeq5 {\rm MeV}$.

We find that 
$\langle L_P \rangle\simeq\langle L_P \rangle_{\rm IR\hbox{-}cut}$  
is numerically satisfied even for each gauge configuration 
in both confined and deconfined phases. 
Table 1 and 2 show the numerical result 
of $\langle L_P \rangle$ and $\langle L_P \rangle_{\rm IR\hbox{-}cut}$ 
in each gauge configuration 
for confined and deconfined phases, respectively.
Thus, the Polyakov loop is almost unchanged by removing 
the contribution from the low-lying Dirac modes \cite{DSI13}, 
which are essential for chiral symmetry breaking.
From both analytical and numerical results, 
we conclude that low-lying Dirac modes 
are not essential modes for confinement, 
which indicates no direct one-to-one correspondence between 
confinement and chiral symmetry breaking in QCD.
\begin{table}[htb]
\caption{
Numerical results for $\langle L_P \rangle$ and 
$\langle L_P \rangle_{\rm IR\hbox{-}cut}$ 
in lattice QCD with $10^3\times5$ and $\beta=5.6$ 
for each gauge configuration, 
where the system is in confined phase.
}
  \begin{tabular}{|c|c|c|c|c|c|c|c|} \hline
configuration No. &1&2&3&4&5&6&7 \\ \hline
Re$\langle L_P \rangle$             &0.00961 &-0.00161&0.0139    &-0.00324&0.000689&0.00423&-0.00807\\ 
Im$\langle L_P \rangle$             &-0.00322&-0.00125&-0.00438&-0.00519&-0.0101  &-0.0168&-0.00265 \\\hline 
Re$\langle L_P \rangle_{\rm IR\hbox{-}cut}$ &0.00961 &-0.00160&0.0139    &-0.00325&0.000706&0.00422&-0.00807 \\ 
Im$\langle L_P \rangle_{\rm IR\hbox{-}cut}$ &-0.00321&-0.00125&-0.00437&-0.00520&-0.0101  &-0.0168&-0.00264 \\\hline
  \end{tabular}
\end{table}
\begin{table}[htb]
\caption{
Numerical results for $\langle L_P \rangle$ and 
$\langle L_P \rangle_{\rm IR\hbox{-}cut}$ 
in lattice QCD with $10^3\times3$ and $\beta=5.7$ 
for each gauge configuration, 
where the system is in deconfined phase.
}
  \begin{tabular}{|c|c|c|c|c|c|c|c|} \hline
configuration No. &1&2&3&4&5&6&7 \\ \hline
Re$\langle L_P \rangle$             &0.316    &0.337     &0.331    &0.305     &0.314   &0.316      &0.337 \\
Im$\langle L_P \rangle$              &-0.00104&-0.00597&0.00723 &-0.00334&0.00167&0.000120&0.0000482 \\\hline
Re$\langle L_P \rangle_{\rm IR\hbox{-}cut}$ &0.319    &0.340     &0.334    &0.307     &0.317   &0.319      &0.340\\
Im$\langle L_P \rangle_{\rm IR\hbox{-}cut}$ &-0.00103&-0.00597&0.00724 &-0.00333&0.00167&0.000121 &0.0000475\\\hline
  \end{tabular}
\end{table}

\section{Summary and concluding remarks}

In this study, we have analytically derived a direct relation between 
the Polyakov loop and the Dirac modes in temporally odd-number lattice QCD 
\cite{SDI13,DSI13}, 
on ordinary square lattices 
with the normal (nontwisted) periodic boundary condition for link-variables.
We have shown that the low-lying Dirac modes have quite small 
contribution to the Polyakov loop \cite{SDI13,DSI13}.

As a new method, we have modified the KS formalism to 
perform the spin-diagonalization of the Dirac operator 
on the temporally odd-number lattice \cite{DSI13}.
In lattice QCD calculations, 
using the ``modified KS formalism'', 
we have numerically shown that 
the contribution of low-lying Dirac modes to the Polyakov loop 
is negligibly small in both confined and deconfined phases \cite{DSI13}.

From the analytical relation (\ref{eq:PLvsD}) and the numerical confirmation, 
we conclude that low-lying Dirac-modes 
have little contribution to the Polyakov loop, 
and are not essential for confinement, 
while these modes are essential for chiral symmetry breaking.
This conclusion indicates no direct one-to-one correspondence between 
confinement and chiral symmetry breaking in QCD.

Since the relation (\ref{eq:PLvsD}) is correct 
in the presence of dynamical quarks 
and also at finite density, 
it is interesting to investigate Eq.(\ref{eq:PLvsD}) 
in full QCD simulations and at finite baryon density.

Our results suggest some independence 
between chiral symmetry breaking and color confinement, 
which may lead to richer phase structure in QCD. 
For example, the phase transition point can be different 
between deconfinement and chiral restoration 
in the presence of strong electro-magnetic fields, 
because of their nontrivial effect on chiral symmetry \cite{ST9193}.

As a future work, it is interesting 
to investigate the Polyakov-loop fluctuation, 
which is recently found to be important 
in the QCD phase transition \cite{LFKRS13}.
It is also meaningful to compare with 
other lattice QCD result on importance of 
infrared gluons to confinement, i.e., confinement originates 
from the low-momentum gluons below 1.5GeV in Landau gauge \cite{YS0809}.

In any case, the research for the direct relation 
between confinement and chiral symmetry breaking 
would give a new direction in the theoretical study of nonperturbative QCD.

\section*{Acknowledgements}
H.S. thanks Prof. E.T. Tomboulis for useful discussions 
and his brief confirmation on our calculations.
H.S. also thanks Profs. J.M. Cornwall and D. Binosi 
for warm hospitality at Trento.
H.S. and T.I. are supported in part by the Grant for Scientific Research 
[(C) No.23540306, Priority Areas ``New Hadrons'' 
E01:21105006, No.21674002] from the Ministry of Education, 
Science and Technology of Japan.  
The lattice QCD calculation was done on NEC-SX8R at Osaka University.

\end{document}